\documentclass{elsart1p}
\usepackage{graphicx}
\usepackage{amssymb}

\begin{document}

\begin{frontmatter}
\title{Charm and charmonium-like spectroscopy in $B$ decays in Belle}

\author{T.A.-Kh.~Aushev\corauthref{cor1}}
\corauth[cor1]{for the Belle Collaboration}
\ead{aushev@itep.ru}
\address{Swiss Federal Institute of Technology of Lausanne, Switzerland\\
Institute for Theoretical and Experimental Physics, Moscow, Russia}

\begin{abstract}
We report measurements of the branching fractions for the decays $B\to
D_{s1}(2536)^+\bar D^{(*)}$, where $\bar D^{(*)}$ is $\bar D^0$, $D^-$
or $D^{*-}$, and study of $B \to X(3872) K$ with $X(3872)$ decaying to
$D^{*0}\bar D^0$ using a sample of 657 million $B\bar{B}$ pairs
recorded at the $\Upsilon(4S)$ resonance with the Belle detector at
the KEKB asymmetric-energy $e^+e^-$ collider.
\end{abstract}

\end{frontmatter}

\section{Introduction}
\label{}

Recent discoveries of a number of new states, such as
$D^*_{s0}(2317)$, $D_{s1}(2460)$, $X(3872)$, $Z(4430)$ and etc., show
that our understanding of charm and charmonium spectroscopy might be
incomplete.  Quark Parton Model can not explain all of these states as
$c\bar s$ or $c\bar c$ mesons.  Some of them could be {\it exotic}
hadrons including tetraquark mesons ($q\bar qq\bar q$) or molecule
states.  Study of the properties of these particles is important for
the understanding of their structures.

The results presented in this article are based on a $605{\rm
  fb}^{-1}$ data sample, corresponding to $657\times 10^6\, B\bar{B}$
pairs, collected at the $\Upsilon(4S)$ resonance with the Belle
detector \cite{belle} at the KEKB asymmetric-energy $e^+ e^-$ collider
\cite{KEKB}.

\section{Study of the decays $B\to D_{s1}(2536)^+\bar D^{(*)}$}
\label{}

The $D_{s1}(2536)^+$ meson is reconstructed in its main decay modes:
$D^{*0}(D^0\pi^0)K^+$, $D^{*0}(D^0\gamma)K^+$ and $D^{*+}K_S$.  The
combinations of $D_{s1}(2536)^+$ and the second charm $D^{(*)}$ meson,
which can be either $\bar D^0$, $D^-$ or $D^{*-}$, with opposite
flavor are considered as $B$ candidates.  Inclusion of charge
conjugate modes is implied throughout the paper.

\begin{figure}[htb]
\begin{center}
  \includegraphics[width=.8\textwidth]{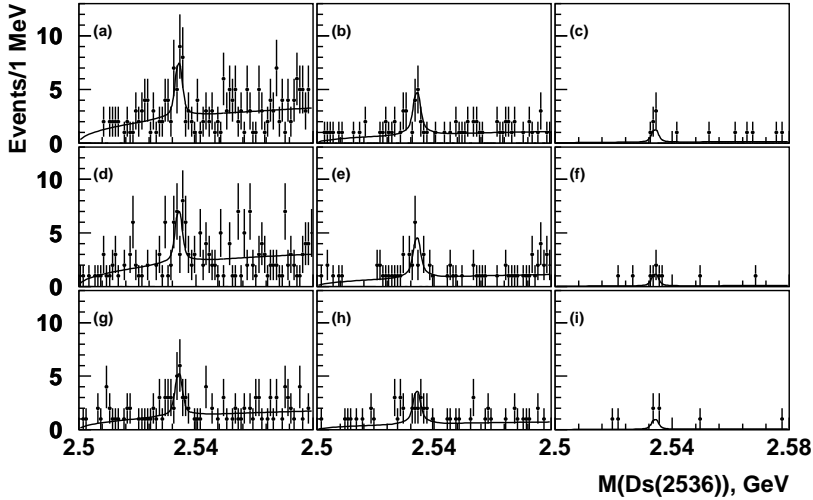}
  \caption{
    $D_{s1}^+(2536)$ mass distributions for the $B$ decays to
    a), b), c) $D_{s1}^+(2536)\bar D^0$,
    d), e), f) $D_{s1}^+(2536) D^-$ and
    g), h), i) $D_{s1}^+(2536) D^{*-}$,
    followed by $D_{s1}^+(2536)$ decays to
    a), d), g) $D_{s1}^+(2536)\to D^{*0}(D^0\gamma)K^+$,
    b), e), h) $D_{s1}^+(2536)\to D^{*0}(D^0\pi^0)K^+$ and
    c), f), i) $D_{s1}^+(2536)\to D^{*+}(D^0\pi^+)K_S$.
    Points with error bars show the data, lines represent the fit result.
  }
  \label{mds2536_data}
\end{center}
\end{figure}

The nine distributions corresponding to three $B$ decay modes times
three $D_{s1}(2536)^+$ decay modes are fitted simultaneously to obtain
the branching fractions for each of the $B$ decay modes
(Fig.~\ref{mds2536_data}).  Statistically significant results are
obtained for each of the decay modes.

\begin{table}[htb]
\begin{center}
  \begin{tabular}{lccc}
    $B$ decay mode \hspace{2cm} & $N$ & \hspace{1cm} ${\cal B}\times10^4$ \hspace{1 cm} & $\cal S$ \\
    \hline
    $D_{s1}(2536)(D^*K)\bar D^0$ & $42.7\pm8.6$ & $3.99\pm0.84\pm0.57$ & $7.0\sigma$\\
    $D_{s1}(2536)(D^*K)D^-$      & $40.4\pm8.7$ & $2.76\pm0.63\pm0.35$ & $6.9\sigma$\\
    $D_{s1}(2536)(D^*K)D^{*-}$   & $33.4\pm7.6$ & $5.03\pm1.21\pm0.68$ & $6.3\sigma$\\
  \end{tabular}
  \caption{Summary of the results for $B\to D_{s1}(2536)^+\bar D^{(*)}$
    decay modes.}
\end{center}
\label{results}
\end{table}

\section{Study of the decay $B\to X(3872) (D^{*0}\bar D^0)K$}

The study described in this paper is a search for the $X(3872)\to
D^{*0}\bar D^0$ decay mode, followed either by $D^{*0}\to D^0\gamma$
or $D^{*0}\to D^0\pi^0$, in charged and neutral $B \to X(3872) K$
decays.  We use the notation $D^{*0}\bar D^0$ to indicate both
$D^{*0}\bar D^0$ and $\bar D^{*0}D^0$.

\begin{figure}[htbp]
  \begin{center}
    \includegraphics[width=0.5\textwidth]{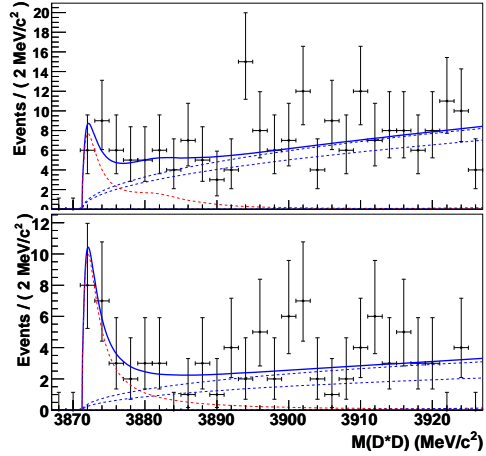}
    \caption{Distribution of $M(D^*D)$ mass for $D^{*0}\to D^0\gamma$
      (top) and $D^{*0}\to D^0\pi^0$ (bottom).  The points with error
      bars are data, the dotted curve is the signal, the dashed curves
      represent the backgrounds.}
    \label{xmass}
  \end{center}
\end{figure}

The $M(D^*D)$ mass distribution is described by a relativistic
Breit-Wigner function convoluted with the mass-dependent Gaussian
resolution for the signal and a square root function for the
background.  This fit gives $50.1^{+14.8}_{-11.1}$ events with a
statistical significance of $8.8\sigma$ (Fig.~\ref{xmass}).  The
branching fraction, averaged over charged and neutral $B$ mesons, is
\begin{eqnarray} 
  \nonumber
  {\cal B}(B \to X(3872) K)\times{\cal B}(X(3872) \to D^{*0}\bar D^0)
  = (0.80 \pm 0.20 \pm 0.11)\times 10^{-4},
\end{eqnarray}
where ${\cal B}(X(3872) \to D^{*0}\bar D^0)$ stands for ${\cal B}(X(3872) \to
D^{*0}\bar D^0) + {\cal B}(X(3872) \to \bar D^{*0}D^0)$.

\section{Conclusion}

In the summary we report the measurement of the branching fractions
for the decays $B\to D_{s1}(2536)^+\bar D^{(*)}$, where $\bar D^{(*)}$
is $\bar D^0$, $D^-$ or $D^{*-}$.  From the simultaneuos fit of all
$D_{s1}(2536)^+$ channels we have measured ${\cal B}(B^+\to
D_{s1}(2536)^+\bar D^0)=(3.99\pm0.84\pm0.57)\times10^{-4}$, ${\cal
  B}(B^0\to D_{s1}(2536)^+ D^-)=(2.76\pm0.63\pm0.35)\times10^{-4}$,
${\cal B}(B^0\to D_{s1}(2536)^+
D^{*-})=(5.03\pm1.21\pm0.68)\times10^{-4}$ with the statistical
significances $7.0\sigma$, $6.9\sigma$ and $6.3\sigma$, respectivly.

We find a near-threshold enhancement in the $D^{*0}\bar D^0$ invariant
mass spectrum at $3872.9^{+0.6+0.3}_{-0.4-0.4}{\rm~MeV}/c^2$ in $B\to
D^{*0}\bar D^0K$ decays.  The significance of this enhancement is
$8.8\sigma$.  The measured branching fraction of the decay is ${\cal
  B}(B \to X(3872) K)\times{\cal B}(X(3872) \to D^{*0}\bar D^0) =
(0.80 \pm 0.20 \pm 0.11)\times 10^{-4}$.

\end{document}